\documentstyle [preprint,revtex]{aps}

\begin{document}

\hfill {MIT-CTP-2169 / ORNL-CCIP-92-15}

\begin{title}
{Kaon-Nucleon Scattering Amplitudes and Z$^*$-Enhancements\\
from Quark Born Diagrams\\
}
\end{title}

\author{T.Barnes}

\begin{instit}
Physics Division and Center for Computationally Intensive Physics\\
Oak Ridge National Laboratory, Oak Ridge, TN 37831-6373\\
and\\
Department of Physics\\
University of Tennessee, Knoxville, TN 37996-1200\\
\end{instit}

\author{E.S.Swanson}

\begin{instit}
Center for Theoretical Physics\\
Laboratory of Nuclear Science and Department of Physics\\
Massachusetts Institute of Technology, Cambridge, MA 02139\\
\end{instit}

\begin{abstract}
We derive closed form
kaon-nucleon scattering amplitudes using the ``quark Born diagram"
formalism, which describes the scattering as a single interaction
(here the OGE spin-spin term)
followed by quark line rearrangement. The low energy
I=0 and I=1 S-wave KN phase shifts
are in reasonably good agreement with experiment given conventional
quark model parameters. For $k_{lab}> 0.7$ Gev however
the I=1 elastic phase shift is larger than predicted by
Gaussian wavefunctions, and we suggest possible reasons for this discrepancy.
Equivalent low energy KN potentials for S-wave scattering are also derived.
Finally we consider OGE forces in the related channels K$\Delta$,
K$^*$N and K$^*\Delta$, and determine which have attractive interactions
and might therefore exhibit strong threshold enhancements or
``Z$^*$-molecule" meson-baryon bound states.
We find that the minimum-spin, minimum-isospin channels
and two additional K$^*\Delta$ channels are
most conducive to the formation of bound states.
Related interesting topics
for future experimental and theoretical studies of
KN interactions are also discussed.
\end{abstract}

\centerline{December 1992}

\newpage

\section{Introduction}

Kaon-nucleon collisions allow one to address many interesting problems
in nuclear and hadron physics \cite{DW}. (By ``kaons" we refer to the
K$^+=u\bar s$ and
K$^0=d\bar s$, generically K, as distinct from the $\bar {\rm K}$ antikaons
K$^-=s\bar u$ and
$\bar {\rm K}^0=s\bar d$.)
Three familiar
examples which we shall discuss below
are 1) the origins of nonresonant ``nuclear"
forces in a system distinct from NN,
2) nuclear structure physics, using kaons as
weakly scattered probes, and
3) searches for possible exotic Z$^*$ baryon resonances which couple
directly to KN. More recently it has become clear that an
understanding of KN scattering in nuclear matter is
important in other areas, such as the interpretation of strangeness production
in nuclear collisions and in two-kaon correlation measurements \cite{had91}.

Elastic KN scattering is a natural
system for the study of nonresonant
nuclear forces.
Since the valence kaon wavefunction contains an $\bar s$ antiquark which
cannot annihilate against the nonstrange nucleon state, direct
production of conventional baryon resonances is excluded.
KN scattering is further simplified by the absence of
one pion exchange, so one can study the nonresonant, non-OPE
part of hadron scattering in relative isolation. Theoretical studies of KN
nuclear forces are especially appropriate because there is already
considerable experimental information on the elastic amplitudes
and two-body inelastic reactions such as KN$\to$ K$^*$N and KN$\to$
K$\Delta$ \cite{DW,bland1,bland2,HARW}.
These experimental amplitudes provide stringent tests for
models of hadronic interactions.
The dominant S-wave elastic
phase shifts are moderately well established, and the higher partial waves
up to L=4 have been determined or estimated \cite{HARW}.
The basic features of the elastic
reaction are a strong repulsion in the I=1 S-wave, a weaker repulsion in the
I=0 S-wave, and an important spin-orbit interaction
which is evident in the P-waves. The important
low energy behavior of the I=0 S-wave,
in particular the scattering length, is unfortunately not yet very well known.
The experimental situation should improve considerably with the development of
new hadronic facilities such as DA$\Phi$NE and KAON
\cite{advert,lf}.

KN scattering also has applications in nuclear
physics; since the kaon-nucleon cross section is relatively small, kaon beams
can be used as probes of nuclear structure. It would obviously be useful to
understand the mechanism and properties of the kaon-nucleon interaction
for this application.
In view of this application one
topic in this paper will be the derivation of effective low energy
KN potentials from the nonrelativistic quark potential model.

Another reason for interest in KN collisions is the possibility of
producing flavor-exotic
Z$^*$ baryon resonances. If discovered, these might be resonances with
the quark valence structure $q^4\bar s$ \cite{mulders},
where $q=$ $u$ or $d$.
Such multiquark hadrons were widely predicted
in the early days of the quark model \cite{multiq},
but it now appears that
multiquark basis states
usually do not support resonances, due to the ``fall apart" effect
\cite{fall,nirev}.
The known exceptions are deuteronlike ``molecule" states of hadron pairs,
which should perhaps be classified as unusual nuclear species.
(Nuclei themselves are excellent examples of the
tendency of multiquark systems to separate into hadronic molecules.)
In the meson-meson
sector two K$\bar {\rm K}$ molecule states are reasonably
well established \cite{WI}, and there are
several other meson-meson candidates \cite{molec}.
In the antikaon-nucleon sector
the $\Lambda(1405)$ is an obvious candidate $\bar {\rm K}$N molecule,
and there presumably are other molecule states in channels with attractive
interactions.
Both the elastic reaction KN$\to $KN and inelastic processes such as KN$\to $
K$^*$N and KN$\to $ K$\Delta$ can be studied for evidence of
exotic Z$^*$ baryon
resonances. With a realistic model
of hadronic interactions we might
reasonably expect to predict the quantum numbers of
exotic meson-baryon molecular
bound states, should these exist.

In this paper we apply the ``quark Born diagram" formalism to KN scattering. In
this approach we assume conventional nonrelativistic quark model wavefunctions
for the asymptotic hadrons, and calculate the Hamiltonian matrix element for
scattering due to a single interaction between constituents in different
incident hadrons.
To form color singlet final states at lowest order one must
then exchange constituents. The full Born amplitude is obtained by summing over
all such processes coherently.
(Similar constituent exchange mechanisms
have been proposed for high energy hadron scattering \cite{CEX}, and there is
strong experimental evidence in favor of this mechanism from large-$t$
exclusive reactions \cite{Baller}.)
This nonrelativistic Hamiltonian matrix element
is then combined with relativistic phase space and kinematics to give results
for differential cross sections, partial wave amplitudes and other scattering
observables. In previous work we derived the elastic scattering amplitudes for
I=2 $\pi\pi$ \cite{BS},
I=3/2 K$\pi$ \cite{BSW}  and I=1 KK \cite{BS}.
(These cases were chosen because they are free of valence $q\bar q$
annihilation processes, which are known to be important if allowed.) We found
good agreement with experimental $\pi\pi$ and K$\pi$ S-wave phase shifts given
conventional quark model parameters. We have also applied similar techniques to
pseudoscalar-vector and vector-vector meson channels \cite{Swan}, and the
results may have important implications for meson spectroscopy \cite{molec}. In
Appendix C of \cite{BS} we presented a diagrammatic representation of these
techniques, with associated ``Feynman rules" for the scattering diagrams. KN
elastic scattering is also annihilation free and affords a nontrivial test of
the quark Born formalism.

KN elastic scattering has previously been the subject of numerous theoretical
investigations. Meson exchange models have been applied in several studies
\cite{mesons}, but these are difficult to justify fundamentally because the
range of heavier meson exchange forces ($\approx 0.2$ fm) is much smaller than
the minimum possible interhadron distance for two distinct hadrons ($\approx 1$
fm) \cite{nirev}.
These models typically have many free parameters, which are not
well established experimentally and are fitted to the data. Thus one is in
effect simply parametrizing experiment. This type of model may be of
theoretical interest as a parametrization of more fundamental scattering
mechanisms which operate at the quark and gluon level, as it may be possible to
relate the predictions of these different approaches.

A quark and gluon approach to scattering using the P-matrix and bag model
wavefunctions was proposed by Jaffe and Low \cite{JL}. They suggested
interpreting the multiquark clusters of the bag model not as resonances, but
instead as the short distance parts of hadron-hadron scattering states. In
principal this approach can be used to predict phase shifts, but in practice it
has mainly been used to interpret experimental phase shifts in terms of
P-matrix poles. This approach has been followed for KN by Roiesnal \cite{Roie},
who concluded that the KN data could indeed be interpreted in terms of poles
approximately at the energies predicted by the bag model, but that the pole
residues (coupling strengths to asymptotic KN channels) did not agree well with
predictions. A more recent bag model calculation of KN scattering by Veit,
Thomas and Jennings \cite{VTJ} used the cloudy bag model, which combines
quark fields (in the baryon) with fundamental pseudoscalar meson fields
in an effective lagrangian. This composite model leads to an I=1
S-wave phase shift and a scattering length
which are very similar to our result, but their I=0 phase shift is much smaller
than experiment. Although this cloudy bag
approach gives promising numerical
results, it does not provide us with an understanding of the scattering
mechanism at the quark and gluon level.

Studies of the dominant S-wave
KN scattering amplitudes in terms of quark model wavefunctions
and quark-gluon interactions
have been published by Bender and Dosch \cite{BD} (adiabatic approach),
Bender, Dosch, Pirner and Kruse \cite{BDPK} (variational generator
coordinate method, GCM) and Campbell and Robson \cite{CR} (resonating group
method, RGM).
The large spin-orbit forces evident in the KN
P-wave data have
also been studied using similar quark model techniques,
first qualitatively by
Pirner and Povh \cite{PP} and later in detail by Mukhopadhyay and Pirner
\cite{MP} (using GCM).
The assumptions regarding dynamics,
the scattering mechanism, quark model wavefunctions
and the parameters used in these calculations are very similar to
our assumptions in this paper.
The most important differences are that 1) our techniques
are perturbative and allow analytic solution, and
2) we disagree about
the size of the OGE contribution to KN scattering. Specifically,
we find that OGE alone suffices to explain the
observed I=1 KN scattering length, whereas Bender {\it et al.} \cite{BDPK}
conclude that OGE is
too small, and that a Pauli blocking effect is dominant in I=1.
Campbell and Robson \cite{CR}
similarly found that the experimental I=1 phase shift
was larger than their predictions, which were based on generalizations of
Gaussian wavefunctions and a full OGE and confining interaction.

\section{Calculation of KN and related scattering amplitudes}

\noindent
{\it a) Hamiltonian and hadron states}
\vskip 0.2cm

Our technique involves a Born order calculation of the matrix element
of the Hamiltonian
between asymptotic hadron states in the nonrelativistic quark model. In the
KN case the dominant interaction was previously found by
Bender {\it et al.} \cite{BDPK} to be the
spin-spin ``color hyperfine" term. A similar conclusion has been
reached for the NN interaction \cite{nirev,hyperf}.
Here we shall adopt
this approximation and neglect the
other OGE and confining terms. Thus, our scattering amplitude is proportional
to the matrix element of
\begin{equation}
H_{scat} = \sum_{a,i<j} \;
\bigg{[}
-{8 \pi \alpha_s\over 3 m_i m_j}\; \delta(\vec r_{ij})
\bigg{]}
\;
\bigg{[}
\vec S_i \cdot \vec S_j
\bigg{]}
\;
\bigg{[}
{\cal F}^a_i \cdot {\cal F}^a_j
\bigg{]}
\end{equation}
\noindent
between asymptotic KN states.
(${\cal F}^a_i$ is the color matrix for quark or antiquark $i$, which is
$\lambda^a/2$ for quarks and $-(\lambda^T)^a/2$ for antiquarks.)
Although we shall quote results for arbitrary
asymptotic hadron wavefunctions, we shall specialize to Gaussian wavefunctions
for our numerical results, as these allow closed form derivation of scattering
observables. Our momentum space Gaussian wavefunctions for the kaon and
nucleon are conventional quark model forms,
\begin{equation}
\phi_{kaon}(\vec p_{rel})
= {1\over \pi^{3/4} \beta^{3/2} }
\exp
\bigg\{
-{\vec p_{rel}^{\; 2} \over 8\beta^2}\; \;
\bigg\}
\end{equation}
where
\begin{equation}
\vec p_{rel} \equiv
{(m_{\bar q}\vec p_q - m_q\vec p_{\bar q})
\over
(m_q + m_{\bar q})/2
}
\ ,
\end{equation}
and
\begin{equation}
\phi_{nucleon}(\vec p_1,\vec p_2,\vec p_3) = {3^{3/4} \over \pi^{3/2}
\alpha^3 }
\exp
\bigg\{
-
{
(
 \vec p_1^{\, 2}
+\vec p_2^{\, 2}
+\vec p_3^{\, 2}
- \vec p_1 \cdot \vec p_2
- \vec p_2 \cdot \vec p_3
- \vec p_3 \cdot \vec p_1
)
\over
3 \alpha^2
}
\bigg\} \ .
\end{equation}
The parameters $\alpha$ and $\beta$ are typically
found to be $\approx 0.3-0.4$ Gev in hadron phenomenology.
These are relative momentum wavefunctions, and have an implicit constraint that
the constituent momenta add to the hadron momentum. In the
full momentum space
wavefunction there is an overall delta function that
imposes this constraint;
\begin{equation}
\Phi_{kaon}(\vec p_q,\vec p_{\bar q};\vec P_{tot}) =
\phi_{kaon}(\vec p_{rel}) \;
\delta(\vec P_{tot} - \vec p_q - \vec p_{\bar q} ) \ ,
\end{equation}
\begin{equation}
\Phi_{nucleon}(\vec p_1,\vec p_2,\vec p_3;\vec P_{tot}) =
\phi_{nucleon}(\vec p_1,\vec p_2,\vec p_3) \;
\delta(\vec P_{tot} - \vec p_1 - \vec p_2 - \vec p_3 ) \ .
\end{equation}
The normalizations are
\begin{displaymath}
\langle \Phi_{kaon}(\vec P'_{tot}) | \Phi_{kaon}(\vec P_{tot}) \rangle
\phantom{yyyyyyyyyyyyyyyyyyyyyyyyyyyyyyy}
\end{displaymath}
\begin{equation}
= \int \!\! \int \!\! \int \!\! \int \,
d\vec p \, d\vec{\bar p} \,
d\vec p\, ' \, d\vec{\bar p }\, ' \,
\Phi^*_{kaon}(\vec p\, ',\vec {\bar p}\, ';\vec P'_{tot})
\Phi_{kaon}(\vec p,\vec {\bar p};\vec P_{tot})
=
\delta(\vec P_{tot} - \vec P'_{tot})
\end{equation}
and
\begin{displaymath}
\langle \Phi_{nucleon}(\vec P'_{tot}) | \Phi_{nucleon}(\vec P_{tot}) \rangle
\phantom{yyyyyyyyyyyyyyyyyyyyyyyyyyyyyyy}
\end{displaymath}
\begin{displaymath}
= \int \!\! \int \!\! \int \!\!  \int \!\!  \int \!\!  \int \,
d\vec p_1 \, d\vec p_2 \, d\vec p_3 \,
d\vec p_1\, ' \, \vec p_2\, ' \, d\vec p_3\, ' \,
\Phi^*_{nucleon}(\vec p_1\, ',\vec p_2\, ',\vec p_3\, ';\vec P'_{tot})
\Phi_{nucleon}(\vec p_1,\vec p_2,\vec p_3;\vec P_{tot})
\end{displaymath}
\begin{equation}
= \delta(\vec P_{tot} - \vec P'_{tot}) \ .
\end{equation}
Since these state normalizations are identical to those used in our previous
study of K$\pi$ scattering \cite{BSW} we can use the relations between
amplitudes and scattering observables given in that reference.

The color wavefunctions for the asymptotic hadrons
are the familiar color singlet states
\begin{equation}
|meson\rangle = \sum_{\imath, \bar \imath=1,3}{1\over \sqrt{3}} \;
\delta_{\imath \bar \imath} \
|\imath \bar \imath\rangle
\end{equation}
and
\begin{equation}
|baryon\rangle = \sum_{i,j,k=1,3}{1\over \sqrt{6}} \;
\epsilon_{ijk} \
|ijk\rangle  \ .
\end{equation}

Our spin-flavor states for the meson and baryon are standard
SU(6) states, but we have found it convenient to write the baryon states
in an unconventional manner, as the usual quark model conventions are
unwieldy for our purposes.
First, to establish our notation, the spin-flavor
K$^+$ state is
\begin{equation}
|{\rm K}^+\rangle =
{1\over \sqrt{2}}
\bigg(
|u_+ \bar s_- \rangle
-
|u_- \bar s_+ \rangle
\bigg) \ .
\end{equation}
For quark model baryon states it is conventional to assign each
quark a fixed location in the state vector, as though identical quarks were
distinguishable fermions. One then explicitly symmetrizes this state.
Thus for example one writes the normalized $\Delta^+(S_z=+3/2)$ state as
\begin{equation}
|\Delta^+(+3/2)\rangle = {1\over \sqrt{3} }
\bigg(
  |u_+ u_+ d_+ \rangle
+ |u_+ d_+ u_+ \rangle
+ |d_+ u_+ u_+ \rangle
\bigg)
\end{equation}
and treats each basis state as orthogonal. Note however that this
is not the usual
way to represent multifermion states. In standard
field theoretic usage each of these basis states is identical to the
others, to within an overall phase. In this language the
normalized $\Delta^+(+3/2)$
state is simply
\begin{equation}
|\Delta^+(+3/2)\rangle = {1\over \sqrt{2} } \;
  |u_+ u_+ d_+ \rangle \ ,
\end{equation}
which we could equally well write as
$ |u_+ d_+ u_+ \rangle /\sqrt{2} $ or
$ |d_+ u_+ u_+ \rangle /\sqrt{2} $. The advantage of using field theory
conventions becomes clear in calculating nucleon matrix elements. For example,
the usual quark model proton state is
\begin{eqnarray}
| P(+1/2) \rangle =&
\bigg( &2|u_+ u_+ d_-  \rangle
-  |u_+ u_- d_+ \rangle
- |u_- u_+ d_+ \rangle
\nonumber \\
&+
&2|u_+ d_- u_+ \rangle  -  |u_+ d_+ u_- \rangle  - |u_- d_+ u_+ \rangle
\nonumber \\
&+
&2|d_- u_+ u_+ \rangle  -  |d_+ u_+ u_- \rangle  - |d_+ u_- u_+ \rangle
\bigg) \bigg{/} \sqrt{18} \ ,
\end{eqnarray}
and in comparison this state in field theory conventions is
\begin{equation}
| P(+1/2) \rangle =
\sqrt{2\over 3}\ \bigg\{ {|u_+ u_+ d_- \rangle \over \sqrt{2}} \bigg\}
-
\sqrt{1\over 3}\ |u_+ u_- d_+ \rangle  \ .
\end{equation}
Use of the latter form, with all permutations of quark entries allowed in
matrix elements, reduces the number of P$\to $P terms from 81 (many of which
are zero) to 4. Of course the results are identical, as these are just
different conventions for the same state.

\vskip 0.2cm
\noindent
{\it b) Enumeration of quark line diagrams for KN}
\vskip 0.2cm

Now we consider KN scattering.
As explained in reference \cite{BS}, we begin by determining the matrix element
of the scattering Hamiltonian (1). First we factor out the overall
momentum conserving delta function and then
derive the remaining matrix element,
which we call
$h_{fi}$;
\begin{equation}
{}_f\langle KN  | H_{scat} | KN  \rangle_i \equiv h_{fi} \ \delta(\vec P_f -
\vec P_i) \ .
\end{equation}

We will discuss one
part of the calculation in detail to explain the techniques, and
then simply quote the full result. Specializing
to the spin up I=1 case K$^+$P(+1/2)$\to$K$^+$P(+1/2),
we require the matrix element
of the scattering hamiltonian (1) between initial and final K$^+$P states
with color and spin-flavor
wavefunctions given by (9,10) and (11,15) respectively.
Since the kaon and proton states (11) and (15)
are each the sum of two terms, the
full amplitude for K$^+$P(+1/2)$\to$K$^+$P(+1/2) is a weighted
sum of 16 subamplitudes.
We shall consider the subamplitude for
$|u_+\bar s_-\rangle \{|u_+ u_+ d_- \rangle / \sqrt{2} \}
\to
|u_+\bar s_-\rangle |u_+ u_- d_+ \rangle $, which we call
$h_{fi}^{e.g.}$, in detail for illustration.

We begin by constructing all allowed quark line diagrams and their
associated combinatoric
factors. First we arrange the initial and final states with their
normalizations on a generic scattering diagram,

\setlength{\unitlength}{2.2pt}
\begin{picture}(200,70)(0,-5)
\put(30,28) {\makebox(0,0)[1]{ $h_{fi}^{e.g.}$ } }
\put(45,28) {\makebox(0,0)[1]{ = } }
\put(180,28) {\makebox(0,0)[1]{(17)} }
\put(65,05) {\makebox(0,0)[1]{${1\over \sqrt{2}}$} }
\put(80,55) {\makebox(0,0)[1]{$u_+ $} }
\put(80,45) {\makebox(0,0)[1]{$\bar s_-$} }
\put(80,15) {\makebox(0,0)[1]{$u_+ $} }
\put(80,05) {\makebox(0,0)[1]{$u_+ $} }
\put(80,-5) {\makebox(0,0)[1]{$d_-$} }
\put(160,55) {\makebox(0,0)[1]{$u_+$} }
\put(160,45) {\makebox(0,0)[1]{$\bar s_-$} }
\put(160,15) {\makebox(0,0)[1]{$u_+$} }
\put(160,05) {\makebox(0,0)[1]{$u_-$} }
\put(160,-5) {\makebox(0,0)[1]{$d_+$} }
\put(170,-5) {\makebox(0,0)[1]{.} }
\put(78,0){
\begin{picture}(75,60)(0,0)
\multiput(5,55)(60,0){2}{\vector(1,0){5}}
\multiput(15,45)(60,0){2}{\vector(-1,0){5}}
\multiput(5,15)(60,0){2}{\vector(1,0){5}}
\multiput(5, 5)(60,0){2}{\vector(1,0){5}}
\multiput(5,-5)(60,0){2}{\vector(1,0){5}}
\put(5,55){\line(1,0){10}}
\put(5,45){\line(1,0){10}}
\put(5,15){\line(1,0){10}}
\put(5,5){\line(1,0){10}}
\put(5,-5){\line(1,0){10}}
\put(65,55){\line(1,0){10}}
\put(65,45){\line(1,0){10}}
\put(65,15){\line(1,0){10}}
\put(65,05){\line(1,0){10}}
\put(65,-5){\line(1,0){10}}
\put(15,-10){\framebox(50,70){}}
\end{picture}
}
\end{picture}

\vskip 1cm
Now we connect the initial and final lines in all possible ways consistent
with flavor conservation.
For the $d$ quark and the $\bar s$ antiquark this choice
is unique. For the final meson's $u$ quark however there are two
choices for which initial baryon's quark it originates from.
Similarly the initial meson's $u$ quark can attach to either of two
final baryon $u$ quarks. Thus we have four quark line diagrams.
We may immediately simplify the diagrams;
since the baryon wavefunctions are symmetric,
we may permute any two initial or final baryon lines and obtain an equivalent
diagram. We use this symmetry to reduce all diagrams to a ``standard form"
in which only the meson's quark and the upper baryon quark are exchanged.
The two choices for the initial baryon's spin up
$u_+$ quark are thus equivalent,
and contribute an overall combinatoric factor of two. The final baryon's quarks
however give inequivalent diagrams, one being nonflip ($u_+(K)\to u_+(P)$)
and the other spin flip ($u_+(K)\to u_-(P)$).
(No polarization selection rules are being imposed yet,
only flavor conservation.)
Thus our amplitude leads to the line diagrams

\vskip 1cm
\setlength{\unitlength}{1.6pt}

\begin{picture}(320,70)(30,-5)
\put(22,28) {\makebox(0,0)[1]{ $h_{fi}^{e.g.}$ } }
\put(31,28) {\makebox(0,0)[1]{ = } }
\put(44,28) {\makebox(0,0)[1]{ ${1\over \sqrt{2}}\cdot 2 \cdot $ }}
\put(160,28) {\makebox(0,0)[1]{ $+$ }}
\put(280,28) {\makebox(0,0)[1]{(18)} }

\put(65,0){

\begin{picture}(75,60)(0,0)

\put(-10,-20){\line(0,1){90}}
\put(-10,-20){\line(1,0){10}}
\put(-10,70){\line(1,0){10}}

\multiput(5,55)(60,0){2}{\vector(1,0){5}}
\multiput(15,45)(60,0){2}{\vector(-1,0){5}}
\multiput(5,15)(60,0){2}{\vector(1,0){5}}
\multiput(5, 5)(60,0){2}{\vector(1,0){5}}
\multiput(5,-5)(60,0){2}{\vector(1,0){5}}
\put(02,54) {\makebox(0,0)[1]{$u_+ $} }
\put(02,44) {\makebox(0,0)[1]{$\bar s_-$} }
\put(02,14) {\makebox(0,0)[1]{$u_+ $} }
\put(02,04) {\makebox(0,0)[1]{$u_+ $} }
\put(02,-6) {\makebox(0,0)[1]{$d_- $} }
\put(80,54) {\makebox(0,0)[1]{$u_+ $} }
\put(80,44) {\makebox(0,0)[1]{$\bar s_-$} }
\put(80,14) {\makebox(0,0)[1]{$u_+ $} }
\put(80,04) {\makebox(0,0)[1]{$u_- $} }
\put(80,-6) {\makebox(0,0)[1]{$d_+ $} }
\put(5,55){\line(1,0){25}}
\put(5,45){\line(1,0){70}}
\put(5,15){\line(1,0){25}}
\put(5,5){\line(1,0){70}}
\put(5,-5){\line(1,0){70}}

\put(50,55){\line(1,0){25}}
\put(50,15){\line(1,0){25}}

\put(30,15){\line(1,2){20}}
\put(30,55){\line(1,-2){20}}



\end{picture}
}

\put(170,0){

\begin{picture}(75,60)(0,0)

\put(90,-20){\line(0,1){90}}
\put(90,-20){\line(-1,0){10}}
\put(90,70){\line(-1,0){10}}

\multiput(5,55)(60,0){2}{\vector(1,0){5}}
\multiput(15,45)(60,0){2}{\vector(-1,0){5}}
\multiput(5,15)(60,0){2}{\vector(1,0){5}}
\multiput(5, 5)(60,0){2}{\vector(1,0){5}}
\multiput(5,-5)(60,0){2}{\vector(1,0){5}}
%
\put(02,54) {\makebox(0,0)[1]{$u_+ $} }
\put(02,44) {\makebox(0,0)[1]{$\bar s_-$} }
\put(02,14) {\makebox(0,0)[1]{$u_+ $} }
\put(02,04) {\makebox(0,0)[1]{$u_+ $} }
\put(02,-6) {\makebox(0,0)[1]{$d_- $} }
\put(80,54) {\makebox(0,0)[1]{$u_+ $} }
\put(80,44) {\makebox(0,0)[1]{$\bar s_-$} }
\put(80,14) {\makebox(0,0)[1]{$u_- $} }
\put(80,04) {\makebox(0,0)[1]{$u_+ $} }
\put(80,-6) {\makebox(0,0)[1]{$d_+ $} }
\put(95,-5) {\makebox(0,0)[1]{.} }
\put(5,55){\line(1,0){25}}
\put(5,45){\line(1,0){70}}
\put(5,15){\line(1,0){25}}
\put(5,5){\line(1,0){70}}
\put(5,-5){\line(1,0){70}}

\put(50,55){\line(1,0){25}}
\put(50,15){\line(1,0){25}}

\put(30,15){\line(1,2){20}}
\put(30,55){\line(1,-2){20}}



\end{picture}
}

\end{picture}

\vskip 1.5cm

We next ``decorate" each of these line diagrams with all possible single
interactions (1) between a quark (or antiquark) in the initial meson and
a quark in the initial baryon. There are six of these per line
diagram (two choices in the meson times three in the baryon),
so we have a total of twelve scattering diagrams to evaluate.
However in this case all but one are trivially zero. Note that
in the first line
diagram we must flip the spins of $u$ and $d$ quarks in the initial
baryon to have a nonzero contribution. This however is not part of our
scattering interaction, which operates between pairs of constituents in
different initial hadrons. The $\vec S_i \cdot \vec S_j$
interaction either flips a pair of spins in different incident hadrons
(through $S_+ S_-$ or $S_-S_+$ terms) or leaves all spins unchanged
(through $S_z S_z$). Thus, the transition in the first line diagram cannot
occur through a single $\vec S_i \cdot \vec S_j$ interaction.
For the second diagram however there is a single nonvanishing transition,
in which the initial meson's $u_+$ quark and the baryon's $d_-$ quark interact
through the spin flip operator;
\vskip 1cm


\setlength{\unitlength}{1.6pt}

\begin{picture}(320,70)(30,-5)
\put(50,28) {\makebox(0,0)[1]{ $h_{fi}^{e.g.}$ } }
\put(70,28) {\makebox(0,0)[1]{ = } }
\put(90,28) {\makebox(0,0)[1]{ $\sqrt{2}\ \ \  \cdot $ }}
\put(280,28) {\makebox(0,0)[1]{(19)} }
\put(230,-5) {\makebox(0,0)[1]{.} }

\put(120,0){

\begin{picture}(75,60)(0,0)

\put(-10,-20){\line(0,1){90}}
\put(-10,-20){\line(1,0){10}}
\put(-10,70){\line(1,0){10}}

\put(90,-20){\line(0,1){90}}
\put(90,-20){\line(-1,0){10}}
\put(90,70){\line(-1,0){10}}

\multiput(5,55)(60,0){2}{\vector(1,0){5}}
\multiput(15,45)(60,0){2}{\vector(-1,0){5}}
\multiput(5,15)(60,0){2}{\vector(1,0){5}}
\multiput(5, 5)(60,0){2}{\vector(1,0){5}}
\multiput(5,-5)(60,0){2}{\vector(1,0){5}}
\put(02,54) {\makebox(0,0)[1]{$u_+ $} }
\put(02,44) {\makebox(0,0)[1]{$\bar s_-$} }
\put(02,14) {\makebox(0,0)[1]{$u_+ $} }
\put(02,04) {\makebox(0,0)[1]{$u_+ $} }
\put(02,-6) {\makebox(0,0)[1]{$d_- $} }
\put(80,54) {\makebox(0,0)[1]{$u_+ $} }
\put(80,44) {\makebox(0,0)[1]{$\bar s_-$} }
\put(80,14) {\makebox(0,0)[1]{$u_- $} }
\put(80,04) {\makebox(0,0)[1]{$u_+ $} }
\put(80,-6) {\makebox(0,0)[1]{$d_+ $} }
\put(5,55){\line(1,0){25}}
\put(5,45){\line(1,0){70}}
\put(5,15){\line(1,0){25}}
\put(5,5){\line(1,0){70}}
\put(5,-5){\line(1,0){70}}

\put(50,55){\line(1,0){25}}
\put(50,15){\line(1,0){25}}

\put(30,15){\line(1,2){20}}
\put(30,55){\line(1,-2){20}}

\put(20,-5){\dashbox{2}(0,60){}}
\multiput(20,-5)(0,60){2}{\circle*{2}}

\end{picture}
}

\end{picture}

\vskip 1.5cm

\noindent
{\it c) Independent quark and gluon diagrams and their
spin and color factors}
\vskip 0.2cm

Finally we require the spin, color, overall phase
and spatial factors associated with this
and the other independent diagrams. There are only four independent
quark and gluon diagrams, since all others can be obtained from these by
permutation of lines. These four diagrams are


\setlength{\unitlength}{1.6pt}

\begin{picture}(320,70)(30,-5)
\put(70,28) {\makebox(0,0)[1]{ $D_1$ } }
\put(100,28) {\makebox(0,0)[1]{ = } }
\put(280,28) {\makebox(0,0)[1]{(20)} }
\put(210,-5) {\makebox(0,0)[1]{,} }

\put(120,0){

\begin{picture}(75,60)(0,0)

\multiput(5,55)(60,0){2}{\vector(1,0){5}}
\multiput(15,45)(60,0){2}{\vector(-1,0){5}}
\multiput(5,15)(60,0){2}{\vector(1,0){5}}
\multiput(5, 5)(60,0){2}{\vector(1,0){5}}
\multiput(5,-5)(60,0){2}{\vector(1,0){5}}
%
\put(5,55){\line(1,0){25}}
\put(5,45){\line(1,0){70}}
\put(5,15){\line(1,0){25}}
\put(5,5){\line(1,0){70}}
\put(5,-5){\line(1,0){70}}

\put(50,55){\line(1,0){25}}
\put(50,15){\line(1,0){25}}

\put(30,15){\line(1,2){20}}
\put(30,55){\line(1,-2){20}}

\put(20,15){\dashbox{2}(0,40){}}
\multiput(20,15)(0,40){2}{\circle*{2}}

\end{picture}
}

\end{picture}

\vskip 1cm


\setlength{\unitlength}{1.6pt}

\begin{picture}(320,70)(30,-5)
\put(70,28) {\makebox(0,0)[1]{ $D_2$ } }
\put(100,28) {\makebox(0,0)[1]{ = } }
\put(280,28) {\makebox(0,0)[1]{(21)} }
\put(210,-5) {\makebox(0,0)[1]{,} }

\put(120,0){

\begin{picture}(75,60)(0,0)

\multiput(5,55)(60,0){2}{\vector(1,0){5}}
\multiput(15,45)(60,0){2}{\vector(-1,0){5}}
\multiput(5,15)(60,0){2}{\vector(1,0){5}}
\multiput(5, 5)(60,0){2}{\vector(1,0){5}}
\multiput(5,-5)(60,0){2}{\vector(1,0){5}}
%
\put(5,55){\line(1,0){25}}
\put(5,45){\line(1,0){70}}
\put(5,15){\line(1,0){25}}
\put(5,5){\line(1,0){70}}
\put(5,-5){\line(1,0){70}}

\put(50,55){\line(1,0){25}}
\put(50,15){\line(1,0){25}}

\put(30,15){\line(1,2){20}}
\put(30,55){\line(1,-2){20}}

\put(20,05){\dashbox{2}(0,50){}}
\multiput(20,05)(0,50){2}{\circle*{2}}

\end{picture}
}

\end{picture}

\vskip 1cm


\setlength{\unitlength}{1.6pt}

\begin{picture}(320,70)(30,-5)
\put(70,28) {\makebox(0,0)[1]{ $D_3$ } }
\put(100,28) {\makebox(0,0)[1]{ = } }
\put(280,28) {\makebox(0,0)[1]{(22)} }
\put(210,-5) {\makebox(0,0)[1]{,} }

\put(120,0){

\begin{picture}(75,60)(0,0)

\multiput(5,55)(60,0){2}{\vector(1,0){5}}
\multiput(15,45)(60,0){2}{\vector(-1,0){5}}
\multiput(5,15)(60,0){2}{\vector(1,0){5}}
\multiput(5, 5)(60,0){2}{\vector(1,0){5}}
\multiput(5,-5)(60,0){2}{\vector(1,0){5}}
%
\put(5,55){\line(1,0){25}}
\put(5,45){\line(1,0){70}}
\put(5,15){\line(1,0){25}}
\put(5,5){\line(1,0){70}}
\put(5,-5){\line(1,0){70}}

\put(50,55){\line(1,0){25}}
\put(50,15){\line(1,0){25}}

\put(30,15){\line(1,2){20}}
\put(30,55){\line(1,-2){20}}

\put(20,15){\dashbox{2}(0,30){}}
\multiput(20,15)(0,30){2}{\circle*{2}}

\end{picture}
}

\end{picture}

\vskip 1cm


\setlength{\unitlength}{1.6pt}

\begin{picture}(320,70)(30,-5)
\put(70,28) {\makebox(0,0)[1]{ $D_4$ } }
\put(100,28) {\makebox(0,0)[1]{ = } }
\put(280,28) {\makebox(0,0)[1]{(23)} }
\put(210,-5) {\makebox(0,0)[1]{.} }

\put(120,0){

\begin{picture}(75,60)(0,0)

\multiput(5,55)(60,0){2}{\vector(1,0){5}}
\multiput(15,45)(60,0){2}{\vector(-1,0){5}}
\multiput(5,15)(60,0){2}{\vector(1,0){5}}
\multiput(5, 5)(60,0){2}{\vector(1,0){5}}
\multiput(5,-5)(60,0){2}{\vector(1,0){5}}
\put(5,55){\line(1,0){25}}
\put(5,45){\line(1,0){70}}
\put(5,15){\line(1,0){25}}
\put(5,5){\line(1,0){70}}
\put(5,-5){\line(1,0){70}}

\put(50,55){\line(1,0){25}}
\put(50,15){\line(1,0){25}}

\put(30,15){\line(1,2){20}}
\put(30,55){\line(1,-2){20}}

\put(20,05){\dashbox{2}(0,40){}}
\multiput(20,05)(0,40){2}{\circle*{2}}

\end{picture}
}

\end{picture}
\vskip 1.5cm
\setcounter{equation}{23}

The spin factor is simply the matrix element of $\vec
S_i \cdot \vec S_j$ for scattering constituents $i$ and $j$,
evaluated between the
initial and final $(q\bar q)(qqq)$ spin states.
This is $1/2$
if both spins $i$ and $j$
are antialigned and both flip, $+1/4$ if the spins are aligned
and neither flips, and $-1/4$ if they are antialigned and neither flips. All
other cases give zero. All spectator spins must not flip or the overall spin
factor is trivially zero.

The color factor can be evaluated using
the states (9), (10) and standard trace techniques, as in (51) of reference
\cite{BS}. The result for each diagram is
\begin{equation}
I_{\rm color}(D_1) = +4/9 \  ,
\end{equation}
\begin{equation}
I_{\rm color}(D_2) = -2/9 \  ,
\end{equation}
\begin{equation}
I_{\rm color}(D_3) = -4/9 \  ,
\end{equation}
\begin{equation}
I_{\rm color}(D_4) = +2/9  \  .
\end{equation}

\vskip 0.2cm
\noindent
{\it d) ``Diagram weights" for KN scattering}
\vskip 0.2cm

We conventionally write the meson-baryon
$h_{fi}$ matrix elements
as row vectors which display the numerical coefficient of each
diagram's spatial overlap integral. Thus,
\begin{equation}
h_{fi} = \bigg{[} \ w_1 ,\ w_2 ,\ w_3 ,\ w_4 \  \bigg{]}
\end{equation}
represents
\begin{equation}
h_{fi} =
  w_1 \, I_{\rm space}(D_1) \;
+ w_2 \, I_{\rm space}(D_2) \;
+ w_3 \, I_{\rm space}(D_3) \;
+ w_4 \, I_{\rm space}(D_4) \ .
\end{equation}
This notation is useful because the diagram weights $\{ w_i \}$
are group theoretic numbers that obey certain symmetries,
whereas the spatial overlap integrals are complicated functions that depend
on the specific spatial wavefunctions rather than the symmetries
of the problem.

As an illustration,
our practice subamplitude $h_{fi}^{e.g.}$ is
\begin{equation}
h_{fi}^{e.g.} = \sqrt{2}\cdot \Big(\, {1\over 2}\, \Big)
\cdot \Big(-{2\over 9}\Big) \cdot
I_{\rm space}(D_2) \  .
\end{equation}
(using the spin and color matrix elements given above), which
we abbreviate as
\begin{equation}
h_{fi}^{e.g.} = \bigg{[} \ 0 ,\ -{\sqrt{2}\over 9} ,\ 0 ,\ 0 \  \bigg{]} \ .
\end{equation}
This completes our detailed derivation
of $h_{fi}^{e.g.}$
for the subprocess
$|u_+\bar s_-\rangle \{|u_+ u_+ d_- \rangle / \sqrt{2} \}
\to
|u_+\bar s_-\rangle |u_+ u_- d_+ \rangle $.

Proceeding similarly, we have derived the weights for the
full KN elastic scattering amplitudes,
given the states (11) and (15) and
their isospin partners.
These are
\begin{equation}
h_{fi}^{\rm KN}({\rm I=0})  =
\bigg{[} \
0 ,\
{1\over 6} ,\
0 ,\
{1\over 6} \  \bigg{]}
\end{equation}
and
\begin{equation}
h_{fi}^{\rm KN}({\rm I=1})  =
\bigg{[} \
{1\over 3}  ,\
{1\over 18} ,\
{1\over 3}  ,\
{1\over 18} \
\bigg{]}
\ .
\end{equation}
For numerical estimates of these amplitudes
we require the spatial overlap integrals, which we shall evaluate
explicitly with Gaussian wavefunctions.

\vskip 0.2cm
\noindent
{\it e) Spatial overlap integrals}
\vskip 0.2cm

The spatial overlap integrals represented by the four diagrams
$D_1\dots D_4$
may be determined using the diagrammatic techniques
discussed in Appendix C of reference \cite{BS}.
These are formally 30-dimensional
overlap integrals (three dimensions times ten external lines), but twelve
integrations are eliminated by external momentum constraints and
an additional nine are eliminated by the unscattered spectator lines.
This leaves a nontrivial 9-dimensional overlap integral for each diagram.
We give the initial meson a label $A$, with three-momentum also called $A$
and quark three-momentum $a$ and
antiquark momentum $\bar a$, and we similarly label the initial baryon $B$, the
final meson $C$ and the final baryon $D$. Since we choose to evaluate these
integrals in the CM frame we use the momentum substitutions $B=-A$ and $D=-C$.
We also introduce a nonstrange to strange quark mass ratio $\rho = m_q / m_s$.
With these substitutions the four spatial overlap integrals are
\begin{displaymath}
I_{space}(D_1) = + {8 \pi \alpha_s \over 3 m_q^2} \; {1\over (2\pi)^3}
\int \!\! \int \!\! \int \, d\vec a \, d\vec b_1 \, d\vec b_2 \
          \phi_A(2a-{2\rho A\over 1+\rho}) \;
          \phi_C^*(2a+{2C\over 1+\rho} - 2A)
\end{displaymath}
\begin{equation}
         \cdot \; \phi_B(b_1, b_2, -A - b_1 - b_2 ) \;
          \phi_D^*(b_1+A-C, b_2, -A - b_1 - b_2 ) \ ,
\end{equation}

\begin{displaymath}
I_{space}(D_2) = + {8 \pi \alpha_s \over 3 m_q^2} \; {1\over (2\pi)^3}
\int \!\! \int \!\! \int \, d\vec b_1 \, d\vec c \, d\vec d_1 \
          \phi_A(2c-{2A\over 1+\rho}  - 2C) \;
          \phi_C^*(2c-{2\rho C\over 1+\rho})
\end{displaymath}
\begin{equation}
         \cdot \; \phi_B(b_1, c, -A - b_1 - c ) \;
          \phi_D^*(d_1,A-C+b_1+c-d_1, -A - b_1 - c) \ ,
\end{equation}

\begin{displaymath}
I_{space}(D_3) = + {8 \pi \alpha_s \over 3 m_q^2}\rho \; {1\over (2\pi)^3}
\int \!\! \int \!\! \int \, d\vec a \, d\vec b_2 \, d\vec c \
          \phi_A(2a-{2\rho A\over 1+\rho}) \;
          \phi_C^*(2c-{2\rho C\over 1+\rho})
\end{displaymath}
\begin{equation}
         \cdot \; \phi_B(a-A+C, b_2, -a - b_2 - C ) \;
          \phi_D^*(a,b_2, -a - b_2 - C) \ ,
\end{equation}

\begin{displaymath}
I_{space}(D_4) = + {8 \pi \alpha_s \over 3 m_q^2}\rho \; {1\over (2\pi)^3}
\int \!\! \int \!\! \int \, d\vec a \, d\vec b_1 \, d\vec c \
          \phi_A(2a-{2\rho A\over 1+\rho}) \;
          \phi_C^*(2c-{2\rho C\over 1+\rho})
\end{displaymath}
\begin{equation}
         \cdot \; \phi_B(b_1, c, -A - b_1 - c ) \;
          \phi_D^*(A-C-a+b_1+c,a, -A - b_1 - c) \ .
\end{equation}
There are many equivalent ways to write these integrals which arise from
different choices of the variables eliminated by momentum constraints.

Note that the overall coefficients of these integrals are positive,
although the
coefficient of $H_{scat}$ (1) is negative. This is because there is an overall
phase factor of $(-1)$ for each diagram $D_1\dots D_4$, due to anticommutation
of quark creation and annihilation operators in the matrix element. Here we
incorporate this phase, which we call the ``signature" of the diagram
\cite{BS}, in the spatial overlap integrals. The signature is equal to
$(-1)^{N_x}$, where $N_x$ is the number of fermion line crossings.
For diagrams $D_1\dots D_4$ above $N_x=3$, so the signature is
\begin{equation}
I_{\rm signature} = (-1) \ .
\end{equation}
Note that a
diagram in nonstandard form, such as the kaon's quark line crossing to the
second baryon quark, can have a $(+1)$ signature;
in the full $h_{fi}$ matrix element this is compensated by a
change in sign of the color factor.

We explicitly evaluate these overlap integrals using the Gaussian wavefunctions
(2) and (4). For Gaussians the integrals
factor into products of three 3-dimensional
integrals, and the results are all of the form
\begin{equation}
I_{\rm space}(D_i) = {8\pi \alpha_s \over 3 m_q^2} {1\over (2\pi)^3}
\; \eta_i \exp \bigg\{ -(A_i - B_i \mu ) P_{cm}^2 \bigg\} \ ,
\end{equation}
where $P_{cm}^2$
is the modulus of each hadron's three-momentum in the CM frame,
$\mu = \cos ( \theta_{CM} ) $ where $\theta_{CM} $ is the
CM scattering angle, and the constants $\eta_i$, $A_i$ and $B_i$ are functions
of $\alpha$, $\beta$ and $\rho$.
$B_i>0$ implies forward peaked scattering and $B_i<0$ is backward peaking.
The pure exponential dependence in
$P_{cm}^2$ and $P_{cm}^2\mu$ is a consequence of the
Gaussian wavefunctions and the
contact interaction.
Introducing the ratio $g = (\alpha / \beta)^2$,
these constants are
\begin{equation}
\eta_1 = 1
\end{equation}
\begin{equation}
A_1 =
{
2\rho^2 + 4\rho + (3g+2)
\over
6 (1+\rho)^2 \alpha^2
}
\end{equation}
\begin{equation}
B_1 =  A_1 \ ,
\end{equation}

\begin{equation}
\eta_2 = \bigg( { 12g \over 7g+6} \bigg)^{3/2}
\end{equation}
\begin{equation}
A_2 =
{
(40g+3)\rho^2 + (32g+6)\rho + (21g^2+28g+3)
\over
6 (7g+6)(1+\rho)^2 \alpha^2
}
\end{equation}
\begin{equation}
B_2 =
{
(-8g+1)\rho^2 + 2\rho + (7g^2+8g+1)
\over
2 (7g+6)(1+\rho)^2 \alpha^2
}
\ ,
\end{equation}
\begin{equation}
\eta_3 = \rho \bigg( { 6 \over g+3} \bigg)^{3/2}
\end{equation}
\begin{equation}
A_3 =
{
(10g+6)\rho^2 + (8g+12)\rho + (7g+6)
\over
6 (g+3)(1+\rho)^2 \alpha^2
}
\end{equation}
\begin{equation}
B_3 =
{
(-g+1)\rho^2 +2\rho + (g + 1)
\over
(g+3)(1+\rho)^2 \alpha^2
}
\ ,
\end{equation}

\begin{equation}
\eta_4 = \rho \bigg( { 12g \over (2g+3)(g+2)} \bigg)^{3/2}
\end{equation}
\begin{equation}
A_4 =
{
(20g^2+40g+3)\rho^2 + (4g^2+14g+6)\rho + (5g^2+10g+3)
\over
6 (2g+3)(g+2)(1+\rho)^2 \alpha^2
}
\ ,
\end{equation}
\begin{equation}
B_4 =
{
(-4g^2-8g+1)\rho^2 + (-4g^2-6g+2)\rho + (g^2+2g+1)
\over
2 (2g+3)(g+2)(1+\rho)^2 \alpha^2
}
\ .
\end{equation}
These results were derived at MIT and ORNL \cite{Mitch} independently using
MAPLE and MACSYMA algebra programs respectively.

Some important properties of these diagrams, specifically
which are forward peaked or backward peaked processes, and which
diagrams dominate at high energies, can be inferred by inspection.
The leading diagram in the high energy limit is $D_1$,
which is a forward peaked exponential in
$t$. The other diagrams are exponentially suppressed in $s$ and are also
forward peaked, with the single exception of $D_4$. Note that for plausible
values of $g\approx 1$ and $\rho\approx 0.6$ this diagram leads to a
{\it backwards} peak ($B_4<0$).
These properties have a simple common origin;
since we are scattering through
a hard delta function interaction, the only angular dependence comes from
overlap suppression due to the spectator lines. A spectator line which is
required to cross into the other hadron gives an especially large suppression
at high energies and small angles. The amplitude for a crossing spectator line
is maximum for backscattered hadrons; in this case the crossing spectator is
actually continuing to move in a new hadron with the same momentum
vector as the
hadron it originally resided in.

The first diagram $D_1$ has no crossing spectators,
so it is not suppressed in $s$;
only the hard scattered constituents
are required to cross into different hadrons. In diagrams $D_2$ and $D_3$
one spectator line is required to cross to a different hadron, so there is
some suppression with increasing $s$. Since {\it two}
spectators do not cross, they dominate the angular dependence, and the
scattering is forward peaked.
Diagram $D_4$, the backward peaking process, is qualitatively different
because two spectator lines are required to change hadrons, and only
one spectator does not cross. In this case ``backwards" meson-baryon scattering
actually corresponds to forward scattering for the
two crossing spectator lines,
which is obviously preferred. This
description attributes
backward peaks, which might otherwise appear counterintuitive, to the
obvious mechanism of ``minimum spectator suppression"
at the quark level.

\vskip 0.2cm
\noindent
{\it f) KN phase shifts and scattering lengths}
\vskip 0.2cm

Given the diagram weights (32-33) and our
results (40-51) for the Gaussian overlap integrals, we have completed the
derivation of the Hamiltonian matrix element $h_{fi}$ for KN elastic
scattering.
Since we have used
the same normalization for KN states as in our previous discussion of K$\pi$
scattering \cite{BSW} we can use the same relations derived there to relate
$h_{fi}$ to scattering variables. First we consider the elastic phase shifts,
which are given by
\begin{equation}
\delta^{KN}_\ell = -{2\pi^2 P_{cm} E_K E_N \over ( E_K + E_N)}
\int_{-1}^1 h_{fi}^{KN}\, P_{\ell} (\mu ) d\mu \ .
\end{equation}
Using the integral
$\int_{-1}^1 e^{b\mu } P_{\ell} (\mu ) d\mu = 2 i_{\ell }(b)$, we find
\begin{equation}
\delta_{\ell }^{KN} = -{4\alpha_s\over 3 m_q^2}
{P_{cm}  E_K E_N \over ( E_K + E_N)} \sum_{i=1}^4
\, w_i \, \eta_i \, \exp ( -A_i P_{cm}^2 ) \; i_\ell (B_i P_{cm}^2 )  \ ,
\end{equation}
where one specifies the isospin state I=0 or I=1 through the choice of
the diagram weights $\{ w_i \} $.

As we approach the KN threshold the
S-wave phase shift is asymptotically linear in $P_{cm}$,
and the coefficient is the scattering length $a_I$. Since the exponential and
the $i_0$ Bessel function are both unity in this limit, we recover a relatively
simple result for the KN scattering length,
\begin{equation}
a^{KN}_I = -{4\alpha_s\over 3 m_q^2}
{M_K M_N \over ( M_K + M_N)} \sum_{i=1}^4
\, w_i \, \eta_i \ .
\end{equation}
Since the coefficients $\{ \eta_i \}$ are relatively simple functions,
we can write these scattering lengths as simple functions of $\alpha_s/m_q^2$,
$\rho = m_q / m_s$, the meson-baryon relative scale parameter
$g = (\alpha / \beta )^2$ and the physical masses $M_K$ and $M_N$.
The results are
\begin{displaymath}
a^{KN}_{I=1} = -{4\alpha_s\over 3 m_q^2}
{M_K M_N \over ( M_K + M_N)}
\phantom{yyyyyyyyyyyyyyyyyyyyyyyyy}
\end{displaymath}
\begin{equation}
\cdot \bigg[
\; {1\over 3} \; + \;
{1\over 18} \bigg( { 12g \over 7g+6}\bigg)^{3/2} \, +
{1\over 3} \; \rho \bigg( {6 \over g+3} \bigg)^{3/2}\, + \,
{1\over 18} \; \rho \bigg( {12g \over (2g+3)(g+2)} \bigg)^{3/2}\
\bigg] \
\end{equation}
and
\begin{displaymath}
a^{KN}_{I=0} =
-{4\alpha_s\over 3 m_q^2}
{M_K M_N \over ( M_K + M_N)}
\phantom{yyyyyyyyyyyyyyyyyyyyyyyyy}
\end{displaymath}
\begin{equation}
\cdot \bigg[
\; {1\over 6} \bigg( { 12g \over 7g+6}\bigg)^{3/2} +
{1\over 6} \; \rho \bigg( {12g \over (2g+3)(g+2)} \bigg)^{3/2}\
\bigg] \ .
\end{equation}
The basic features of the low energy KN interaction, a repulsive I=1 S-wave
and a repulsive but less strong I=0 S-wave, are already evident in these
formulas. (The parameter $g = (\alpha/\beta)^2$ is
constrained by quark model phenomenology to be comparable to
unity.)
Detailed numerical results for the scattering lengths and phase shifts and a
comparison with experiment are presented in the next section.

\section{Comparison with experiment}

\noindent
{\it a) Scattering lengths}
\vskip 0.2cm

Before we discuss our numerical predictions we first review the status of the
experimental scattering lengths. Since there are unresolved disagreements
between analyses in the I=0 channel, we have compiled relatively recent
(since 1980) single-energy S-wave phase shifts for our discussion. These are in
chronological order Martin and Oades \cite{MO} (Aarhus and UC London, 1980);
Watts {\it et al.} \cite{Watts} (QMC and RAL, 1980); Hashimoto \cite{Hash}
(Kyoto and VPI, 1984); and Hyslop {\it et al.} \cite{HARW} (VPI, 1992). The I=1
data set analysed by Arndt and Roper \cite{AR} (VPI, 1985) was incorporated in
the 1992 VPI simultaneous analysis of I=0 and I=1 data, so we shall not
consider it separately. The energy dependent parametrizations of Corden {\it et
al.} \cite{Corden} and Nakajima {\it et al.} \cite{Nakajima} are not included
in our discussion.

In Fig.1 we show these experimental I=0 and I=1 S-wave phase shifts versus
$P_{cm}=|\vec P_{cm}|$. The linear low energy behavior which determines the
scattering length is evident in the I=1 data, and Hyslop {\it et al.} cite a
fitted value of $a^{KN}_{I=1} = -0.33$ fm. Previous analyses (summarized in
\cite{DW} and \cite{HARW}) have given values between $-0.28(6)$ fm
\cite{Cutkosky} and $-0.33$ fm \cite{HARW,Hyslop}. A more useful way to present
the S-wave phase shift data is to display $\delta_0^I / P_{cm}$ versus
$P_{cm}^2$; the intercept is the scattering length, and the slope at intercept
determines the effective range. In Fig.2 we show the
S-wave phase shifts
in this manner; an I=1 scattering length of about
$-0.31(1)$ fm is indeed evident, which we shall take as our estimated
experimental value.

Unfortunately the I=0 scattering length is much less well determined, as is
evident in Figs.1 and 2.
Previous (favored) solutions up to 1982 are summarized in Table
2.3 of \cite{DW}, and range between $+0.02$ fm and $-0.11^{+0.06}_{-0.04}$ fm.
There appear to be two sets of low energy values in the data of Fig.1, a
smaller phase shift from the Aarhus-UCL and QMC-RAL collaborations and a larger
one from from the Kyoto-VPI and VPI analyses. Below $P_{cm}=0.4$ Gev the
Kyoto-VPI and VPI results are larger than Aarhus-UCL and QMC-RAL
by about a factor of two. The VPI group actually
cite a scattering length of $a^{KN}_{I=0} = 0.0$ fm, although this requires
rapid low energy variation below the first experimental point (compare their
Fig.1(a) with the I=1 phase shift in their Fig.2(a), which is constrained by
experiment at lower energy and shows the expected $\sqrt{T_{lab}}\propto
P_{cm}$ S-wave dependence). Since the I=1 phase shift is close to linear for
$P_{cm}< 0.4$ Gev  ($k_{lab}< 0.7$ Gev), we will assume that the zero I=0
scattering length quoted in \cite{HARW} is an artifact of their fit, and that
the actual I=0 phase shift is approximately linear in $P_{cm}$ for $P_{cm}<
0.4$ Gev. We can then read the I=0 scattering length from the
intercept in Fig.2. From the figure we see that a naive extrapolation to
threshold leads to scattering lengths of about $-0.09(1)$ fm and $-0.17(2)$
fm respectively from the two sets of references. In summary, the experimental
phase shifts shown in Fig.2 suggest to us the scattering lengths

\begin{eqnarray}
 &a^{KN}_{I=1}(expt.)\phantom{\bigg|_{Aarhus-QMC-RAL-UCL}}
&=  -0.31(1) \ {\rm fm} \ ; \nonumber \\
 &a^{KN}_{I=0}(expt.)\bigg|_{Aarhus-QMC-RAL-UCL} &= -0.09(1) \ {\rm fm} \ ,
\nonumber \\
 &a^{KN}_{I=0}(expt.)\bigg|_{Kyoto-VPI}^{\phantom{Aarhus-QMC-RAL-UCL}}
&=  -0.17(2) \ {\rm fm} \ .
\end{eqnarray}
We emphasize that the I=0 values are our
interpretation of the data from Fig.2, and the references cited
quote smaller scattering lengths that we believe the data does not support.
As the values of the I=0 scattering length and low energy phase shifts
are controversial, an accurate determination should be a first priority at a
kaon facility.

To compare our predictions with experiment we first use a ``reference parameter
set" with conventional quark model parameters. The hyperfine strength is taken
to be $\alpha_s / m_q^2 =0.6/(0.33)^2$ Gev$^{-2}$, and the nonstrange to
strange quark mass ratio is $\rho = m_q/m_s= 0.33\; {\rm Gev}\; / 0.55\; {\rm
Gev} = 0.6$. The remaining parameter in the scattering length formulas is
$g=(\alpha/\beta)^2$, the ratio of baryon to meson width parameters squared.
These parameters are rather less well determined phenomenologically. For
baryons, values in the range $\alpha=0.25-0.41$ Gev have been used in
nonrelativistic quark model studies \cite{Simon,Nathan,Roman}.
Isgur and Karl \cite{IKK} originally used $\alpha=0.32$ Gev for
spectroscopy, but Copley, Karl and Obryk \cite{CKO} had earlier found that the
photocouplings of baryon resonances required a somewhat larger value of
$\alpha=0.41$ Gev, which may be a more realistic estimate \cite{Simon,Nathan}
because it is less sensitive to short distance hyperfine matrix elements.
This larger value was also found
by Koniuk and Isgur \cite{KI} for baryon electromagnetic transition amplitudes.
Here
we take $\alpha=0.4$ Gev as our reference value. For mesons, studies of various
matrix elements have led to values of $\beta=0.2-0.4$ Gev \cite{Nathan}. In our
previous study of I=2 $\pi\pi$ scattering we found a best fit to the S-wave
phase shift data with $\beta=0.337$ Gev. Here we use a similar $\beta=0.35$ Gev
as our reference value; if the quark Born formalism is realistic we should use
essentially the same meson parameters in all reactions.

With our reference parameter set and physical masses $M_K=0.495$ Gev and
$M_N=0.940$ Gev, our formulas (55) and (56) give
\vskip 0.5cm
\begin{equation}
a^{KN}_{I=1}(ref.\; set) = -0.35 \ {\rm fm}
\end{equation}
and
\begin{equation}
a^{KN}_{I=0}(ref.\; set) = -0.12 \ {\rm fm} \ .
\end{equation}
\vskip 0.5cm
In view of our approximations, the parameter uncertainties, and the
uncertainties in the I=0 data, these scattering lengths compare rather well
with experiment. Note that our conclusions differ from those of Bender {\it et
al.} \cite{BDPK}, who reported that the OGE contribution to I=1 scattering was
too small to explain the observed S-wave phase shift. We discuss this
disagreement further in the appendix.

Now suppose we attempt to fit our estimated
experimental values of the scattering lengths (57)
by varying our parameter set. It is useful to fit the ratio
$a^{KN}_{I=0}/a^{KN}_{I=1}$, since this involves only $\rho$ and the width
parameter $g$. We have fixed $\rho=0.6$, and in any case we find that
$a^{KN}_{I=0}/a^{KN}_{I=1}$ is insensitive to $\rho$, so only $g$ remains as an
important parameter. In Fig.3 we show the predicted ratio of KN scattering
lengths as a function of $\alpha/\beta$. The two experimental ratios
assuming the values in (57) are also indicated. The larger ratio
$a^{KN}_{I=0}/a^{KN}_{I=1} = 0.17/0.31$ requires $\alpha/\beta = 1.91$,
rather far from typical quark model values. Fitting the smaller ratio
$a^{KN}_{I=0}/a^{KN}_{I=1} = 0.09/0.31$ requires $\alpha/\beta = 1.02$, which
is more representative of quark model parameters. An accurate determination of
the I=0 KN scattering length through direct low energy measurements, rather
than by extrapolation, would be a very useful experimental contribution; this
would allow a more confident test of our results and those of other models
(as shown for example in Table 6-4 of Hyslop \cite{Hyslop}).

\vskip 0.2cm
\noindent
{\it b) S-wave phase shifts}
\vskip 0.2cm

The S-wave KN phase shifts
predicted by (53) with $\ell =0$ given the ``reference parameter set"
$\alpha_s/m_q^2 = 0.6/(0.33)^2$ Gev$^{-2}$, $\rho=m_q/m_s=0.6$,
$\alpha=0.4$ Gev and
$\beta=0.35$ Gev are shown as dashed lines
in Fig.4. As we noted in the previous section,
this parameter set gives reasonable scattering lengths, although
the I=0 scattering length is not yet very well established experimentally.

At higher energies the reference parameter set predicts an I=1 phase shift
that retreats more quickly with energy than is observed experimentally;
in Fig.4 we see a rapid departure of theory and experiment above
$P_{cm}=0.4$ Gev ($k_{lab}=0.7$ Gev).
This is near the opening of the inelastic channels
K$\Delta$ and K$^*$N,
as indicated in Fig.4.
Two possible reasons for
this discrepancy are 1) inelastic effects of the channels K$\Delta$, K$^*$N
and K$^*\Delta$, which should become important just where theory and
experiment part, and 2) short distance components in the meson and baryon
wavefunctions that are underestimated by the smooth Gaussian
wavefunctions (2) and (4).

Although inelastic effects are certainly important
experimentally \cite{bland1,bland2},
the most important low energy inelastic process
is P-wave K$\Delta$ production \cite{bland2}.
Hyslop {\it et al.} \cite{HARW} similarly find relatively small inelasticities
in the I=1 KN S-wave, with $\eta \geq 0.9$ for $P_{cm}\leq 0.68$ Gev. At the
end of this range our predicted phase shift
given the reference parameter set is only about half the observed
value, so it appears unlikely that the discrepancy is mainly due to inelastic
channels.

A second possible reason for the discrepancy is a departure of the hadron
wavefunctions from the assumed single Gaussian forms at short distances; both
the meson $q\bar q$ states and the baryon $qq$ substates experience attractive
short distance interactions from the color Coulomb and hyperfine terms (for
spin singlets), which will lead to enhancements of the short distance
components of their wavefunctions and increased high energy scattering
amplitudes. If this is the principal reason for the discrepancy, we would
expect a global
fit to the S-wave phase shifts to prefer a smaller
hadron length scale. In Fig.4 we show the result of a three-parameter fit to
the full 1992 VPI I=0,1 energy independent S-wave data set \cite{HARW}, letting
$\alpha_s/m_q^2, \alpha$ and $\beta$ vary and holding $\rho=0.6$ fixed. The fit
is shown as solid lines, and is evidently quite reasonable both near threshold
and at higher energies. The fitted parameters are
$\alpha_s/m_q^2=0.59/(0.33)^2$ Gev$^{-2}$, $\alpha=0.68$ Gev and $\beta=0.43$
Gev; the hyperfine strength is a typical quark model result but the width
parameters $\alpha$ and $\beta$ are about 1.5 times the usual quark model
values. Thus, a fit to the S-wave VPI data using single Gaussian wavefunctions
requires a hadron length scale about 0.7 times the usual scale. This result is
largely independent of the data set chosen, since it is driven by
the large I=1 phase shift, which shows little variation between
analyses. Evidently the predicted S-wave phase shifts at higher energies are
indeed very sensitive to the short distance parts of the wavefunction; this
supports our conjecture that the discrepancy at higher energies is
an artifact of our single Gaussian wavefunctions.
A calculation of these S-wave phase shifts using realistic Coulomb
plus linear plus hyperfine wavefunctions is planned \cite{Simon} and
should provide a very interesting test of
the quark Born formalism.

\vskip 0.2cm
\noindent
{\it c) Higher-L partial waves, spin-orbit and inelastic effects}
\vskip 0.2cm

In addition to the S-wave phase shifts, higher-L KN elastic phase shifts
and properties of the inelastic reactions KN$\to$K$^*$N, KN$\to$K$\Delta$ and
KN$\to$K$^*\Delta$ have been the subjects of experimental investigations.
These studies have found important effects in the L$>$0 partial waves
which are beyond the scope of the present paper.

One especially interesting effect is a remarkably large spin-orbit interaction
in the I=0 KN system; the L=1 states have a large, positive phase shift for
J=1/2 and a weaker, negative phase shift for J=3/2 (see \cite{HARW} and
references cited therein). This spin-orbit interaction cannot arise in our
quark Born amplitudes given the approximations we have made in this paper;
since we have incorporated only the spin-spin hyperfine interaction in single
hadronic channels, our phase shifts (53) are functions of the total hadronic L
and S but not J. Some but not all of this spin-orbit interaction may simply
require incorporation of the OGE spin-orbit term; Mukhopadhyay and Pirner
\cite{MP} found that the quark spin-orbit interaction was sufficient to
explain the sign and magnitude of some of the weaker KN spin-orbit forces, but
that the I=0, J=1/2 phase shift was much too large to be explained as an OGE
force. The strong KN spin-orbit forces might conceivably be due to couplings to
inelastic channels; since the available mixing states and their couplings to KN
are J-dependent, they might lead to effective spin-orbit forces at the hadronic
level, even if we do not include spin-orbit forces at the quark level. We hope
to treat this interesting possibility in a future study of coupled channel
effects using the quark Born formalism.

Since we do not have a model of the large spin-orbit effect it is not
appropriate to include a detailed discussion of our predicted amplitudes and
cross sections at higher energies, where higher partial waves are important.
In the interest of completeness however we will briefly discuss our
predicted differential cross section at high energy, since we previously noted
that we found an exponential in $t$ in I=2 $\pi\pi$ scattering, reminiscent of
diffraction in magnitude but not in phase \cite{BS}.
The differential cross section in this unequal mass case is related to the
$h_{fi}$ matrix element (29) by
\begin{equation}
{d\sigma \over dt} = 4\pi^5 \,
{ \Big[ \, s^2 - (M_N^2 - M_K^2)^2 \, \Big]^2 \over
s^2 \, \Big[ \, (s-(M_N+M_K)^2) (s-(M_N-M_K)^2) \,  \Big] } \; | h_{fi} |^2 \ .
\end{equation}
For KN scattering in the
high energy limit only the contribution from diagram $D_1$ (20) survives, and
we find
\begin{equation}
\lim_{s\to\infty} {d\sigma \over dt} = {4\pi\alpha_s^2\over 9m_q^4} \,
w_1^2 \, \exp \{ A_1 t \} \ .
\end{equation}
Thus we again find an exponential in $t$ at high energy, with a slope parameter
(41) that is numerically equal to
\begin{equation}
b = A_1 = 3.7 \; {\rm Gev}^{-2}
\end{equation}
given our reference parameter set. This is similar to the observed
diffractive I=1 KN slope parameter \cite{Carnegie} of
\begin{equation}
b(expt.) \approx 5.5-5.9 \; {\rm Gev}^{-2} \ .
\end{equation}
The normalizations of the theoretical and experimental I=1 high energy
differential cross sections however differ by about an order of magnitude, and
are 1.8 mb Gev$^{-2}$ (reference parameter set) versus $\approx 15$ mb
Gev$^{-2}$ (experiment \cite{Carnegie}). We noted a similar tendency for the
reference parameter set to underestimate high energy amplitudes in our
discussion of the S-wave phase shifts, which we attributed to the
single Gaussian wavefunction approximation. One interesting prediction is that
I=0 KN scattering should have no diffractive peak in the high energy limit,
since it has has $w_1=0$; unfortunately there is no I=0 high energy data to
compare this prediction with. A serious comparison with high energy scattering
will presumably require the use of wavefunctions with more realistic
high momentum components as well as the incorporation of inelastic channels,
which may strongly affect the elastic amplitudes.

\section{KN equivalent potentials}

Sufficiently close to threshold our quark Born scattering amplitudes can be
approximated by local potentials. These potentials are useful in applications
such as multichannel scattering and investigations of possible bound states,
which are easiest to model using a Schr\"odinger equation formalism with local
potentials. There are many ways to define an equivalent low energy potential
from a scattering amplitude such as $h_{fi}$; several such procedures
are discussed in \cite{Swan,BG} and in Appendix E of \cite{BS}. Of course
effective potentials extracted using different definitions can appear to
be very different functions of $r$ although they lead to similar low energy
scattering amplitudes.

One approach to defining an equivalent potential is to derive a potential
operator $V_{op}(r)$ which give the scattering amplitude $h_{fi}$ in Born
approximation. This ``Born-equivalent potential" technique is discussed in
reference \cite{BG} and in Appendix E of \cite{BS}; it has been tested on the
OGE interaction, from which one recovers the correct Breit-Fermi Hamiltonian at
$O(v^2/c^2)$ \cite{BG}. To derive the Born-equivalent potential we reexpress
our scattering amplitude in the CM frame as a function of the transferred
three-momentum $\vec q = \vec C - \vec A$ and an orthogonal variable $\vec
{\cal P} = (\vec A + \vec C)/2$. We then expand the scattering amplitude in a
power series in $\vec {\cal P}$ and equate the expansion to the Born expression
for nonrelativistic potential scattering through a general potential operator
$V_{op}(r)$, which may contain gradient operators. The leading term, of order
${\cal P}^0$, gives the Born-equivalent local potential $V(r)$.

In this meson-baryon scattering problem our Hamiltonian matrix elements are of
the form
\begin{equation}
h_{fi} = {8\pi \alpha_s \over 3 m_q^2} {1\over (2\pi)^3} \sum_{i=1}^4 w_i
\eta_i \exp \bigg\{ -(A_i - B_i \mu ) P_{cm}^2 \bigg\} \ .
\end{equation}
Making the required substitutions
$P_{cm}^2 = \vec {\cal P}^2 + \vec q^{\, 2}/4$ and
$P_{cm}^2\mu = \vec {\cal P}^2 - \vec q^{\, 2}/4$ and Fourier transforming with
respect to $\vec q$ as in \cite{BS} gives the equivalent low energy KN
potential
\begin{equation}
V_{KN}(r) = {8  \alpha_s \over 3 \sqrt{\pi} m_q^2}\ \sum_{i=1}^4
{w_i \eta_i \over ( A_i + B_i )^{3/2} }  \;
\exp\bigg\{ - r^2/ (A_i + B_i) \bigg\} \ .
\end{equation}
Thus our Born-equivalent meson-baryon potentials are sums of four Gaussians,
one from each inequivalent quark Born diagram, weighted by the diagram weights
of that channel.

The potentials for I=0 and I=1 with our reference parameter set $\alpha_s=0.6$,
$m_q=0.33$ Gev, $\rho=m_q/m_s=0.6$, $\alpha=0.40$ Gev and $\beta=0.35$ Gev are
shown in Fig.5. They are repulsive and have a range of about 0.3 fm, as one
would expect for a short range ``nuclear" core. The potentials at contact are
rather similar in this formalism, and the relative weakness of I=0 scattering
is a result of its shorter range. This is an effect of the backward peaking
diagram $D_4$, which leads to a very short range potential with a large value
at contact, and carries higher weight in I=0 scattering.

Although these Born-equivalent potentials are convenient for use in a
meson-baryon Schr\"odinger equation, the actual KN potentials are so strong
that they reproduce some features of the interaction only qualitatively. For
example, the Born diagrams give an I=1 scattering length of $-0.35$ fm, but the
Born-equivalent potential (65) for I=1 in the Schr\"odinger equation for KN
leads to a scattering length of only about $-0.22$ fm. The discrepancy is due
to higher order effects of $V_{KN}(r)$ in the Schr\"odinger equation; we have
confirmed that the ratio of $h_{fi}$ and $V_{KN}(r)$ scattering lengths
approaches unity in the small-$\alpha_s$ limit. In a multichannel study one
might modify $V_{KN}(r)$ (65) to give the input $h_{fi}$ scattering lengths,
perhaps through a change in the overall normalization, as a way of providing a
more realistic potential model of the quark Born amplitudes.

\section{Results for K$\Delta$, K$^*$N and K$^*\Delta$; prospects for
Z$^*$-molecules}

The channels K$\Delta$, K$^*$N and K$^*\Delta$ are interesting in part because
they may support molecular bound states if the effective interaction is
sufficiently attractive. In contrast the low energy KN interaction is
repulsive in both isospin states. These ``Z$^*$-molecules" would appear
experimentally as resonances with masses somewhat below the thresholds of
$\approx 1.7$ Gev, $\approx 1.85$ Gev and $\approx 2.1$ Gev. Even if there
are no bound states, attractive interactions will lead to threshold
enhancements which might be misidentified as Z$^*$ resonances just above
threshold.

Plausible binding energies of hadronic molecules can be estimated from the
uncertainty principle and the minimum separation allowed for distinct hadrons
as $E_B \sim 1/ ( M_{had}\cdot 1\, {\rm fm}^2) \sim 50$ Mev. In comparison, the
best established molecules or molecule candidates have binding energies ranging
from 2.2 Mev (the deuteron, which has a repulsive core) through 10-30 Mev (the
$f_0(975)$, $a_0(980)$ and $\Lambda(1405)$). (The $f_0(1710)$, with a binding
energy relative to K$^*\bar {\rm K}^*$ of about 75 Mev, appears plausible but
is a more controversial candidate \cite{molec}.) Finally, the state $f_2(1520)$
seen by the Asterix \cite{Asterix}, Crystal Barrel \cite{CB} and Obelix
\cite{Obelix} collaborations in P$\bar {\rm P}$ annihilation is an obvious
candidate for a nonstrange vector-vector molecule, with a (poorly determined)
binding energy relative to $\rho\rho$ threshold of perhaps 20 Mev.

Several candidate Z$^*$ resonances which might be meson-baryon molecule states
have been reported in KN partial wave analyses. The 1986 Particle Data Group
compilation \cite{PDG86} (the most recent to review the subject of Z$^*$
resonances) cited two I=0 candidates,
[$Z_0(1780), {1\over 2 }^+$] and
[$Z_0(1865), {3\over 2 }^-$] and four I=1 possibilities,
[$Z_1(1725), {1\over 2 }^+$];
[$Z_1(1900), {3\over 2 }^+$];
$Z_1(2150)$ and
$Z_1(2500)$. However the evidence for these states is not strong, and the
PDG argue that the standards of proof must be strict in this exotic channel.
For this reason these states were only given a one star ``Evidence weak; could
disappear." status. The 1986 PDG also noted that ``The general prejudice
against baryons not made of three quarks and the lack of any experimental
activity in this area make it likely that it will be another 15 years before
the issue is decided.". The 1992 PDG compilation \cite{PDG92} makes a similar
statement, with ``15 years" revised to ``20 years".

In their recent analysis of the data Hyslop {\it et al.} \cite{HARW} summarize
some previous claims and report evidence for ``resonancelike structures"
[$Z_0(1831),{1\over 2}^+$];
[$Z_0(1788),{3\over 2}^-$];
[$Z_1(1811),{3\over 2}^+$] and
[$Z_1(2074),{5\over 2}^-$]. The negative parity candidates
$Z_0(1788)$ and
$Z_1(2074)$ have
quantum numbers and masses consistent with S-wave K$^*$N and K$^*\Delta$
molecules respectively. We would not normally expect P-wave molecules; odd-L is
required to couple to positive parity KN channels, and the centrifical barrier
suppresses binding due to these short range forces. However, threshold effects
which resemble resonances might arise in the full multichannel problem, and the
very strong spin-orbit force evident in the P$_{01}$ and P$_{03}$ KN partial
waves may be sufficient to induce binding in some channels.
A clarification of the status of Z$^*$ candidates
through the determination of experimental amplitudes for the processes KN $\to
$ K$^*$N, KN $\to $ K$\Delta$ and KN $\to $ K$^*\Delta$ in addition to the
elastic KN reaction will be an important goal of future studies at kaon
factories.

All the S-wave (I,J$^{\rm P}$) quantum numbers, in which
molecule bound states are {\it a priori} most likely, are as follows;
\begin{displaymath}
{\rm K}\Delta (\approx 1.6-1.7 \ \hbox{Gev}): \ \ \
(2,{3\over 2}^-) \ ;
(1,{3\over 2}^-) \ ; \\
\end{displaymath}

\begin{displaymath}
{\rm K}^* {\rm N} (\approx 1.75-1.85 \ \hbox{Gev}): \ \ \
(1,{3\over 2}^-) \ ;
(1,{1\over 2}^-) \ ; \\
(0,{3\over 2}^-) \ ;
(0,{1\over 2}^-) \ ;
\end{displaymath}

\begin{displaymath}
{\rm K}^* \Delta (\approx 2.0-2.1 \ \hbox{Gev}): \ \ \
(2,{5\over 2}^-) \ ;
(2,{3\over 2}^-) \ ;
(2,{1\over 2}^-) \ ; \\
(1,{5\over 2}^-) \ ;
(1,{3\over 2}^-) \ ;
(1,{1\over 2}^-) \ .
\end{displaymath}

We can use our detailed model of meson-baryon scattering in the $(q\bar s)
(qqq)$ system $(q=u,d)$ to identify channels which experience attractive
interactions as a result of the color hyperfine term. These we again show as
weight factors which multiply each of the four diagrams $D_1\dots D_4$. Since
the overlap integrals these weights multiply are all positive and of comparable
magnitude, the summed weight can be used as an estimate of the sign and
relative strength of the interaction in each channel. Positive weights
correspond to a repulsive interaction. Our results for the $h_{fi}$ ``diagram
weights" for all K$\Delta$, K$^*$N and K$^*\Delta$ channels in (I,S$_{tot}$)
notation are given below. We also give the numerical values we find for the
scattering length in each channel given our reference parameter set and masses
M$_{K^*}=0.895$ Gev and M$_{\Delta}=1.210$ Gev.
\vskip 0.5cm

\begin{equation}
{{\rm K}\Delta}(2,{3\over 2})  =
{1\over 6} \ \bigg{[} \
+{3}  ,\
-{1} ,\
+{3}  ,\
-{1} \
\bigg{]}
\ ;
\end{equation}
\begin{equation}
a = -0.38 \ \hbox{ fm} \ .
\end{equation}
\vskip 0.5cm
\begin{equation}
{{\rm K}\Delta}(1,{3\over 2})  = -{1\over 3} \
{{\rm K}\Delta}(2,{3\over 2})  =
{1\over 18} \ \bigg{[} \
-{3}  ,\
+{1} ,\
-{3}  ,\
+{1} \
\bigg{]}
\ ;
\end{equation}
\begin{equation}
a = +0.13 \ \hbox{ fm} \ .
\end{equation}
\vskip 0.5cm
\begin{equation}
{{\rm K}^*{\rm N} }(1,{3\over 2})  =
{1\over 27} \ \bigg{[} \
+{7}  ,\
+{1} ,\
-{5}  ,\
 {0} \
\bigg{]}
\ ;
\end{equation}
\begin{equation}
a = -0.08 \ \hbox{ fm} \ .
\end{equation}
\begin{equation}
{{\rm K}^*{\rm N} }(1,{1\over 2})  =
{1\over 54} \ \bigg{[} \
+26  ,\
+5 ,\
+2  ,\
-3 \
\bigg{]}
\ ;
\end{equation}
\begin{equation}
a = -0.39 \ \hbox{ fm} \ .
\end{equation}
\vskip 0.5cm
\begin{equation}
{{\rm K}^*{\rm N} }(0,{3\over 2})  =
{1\over 9} \ \bigg{[} \
+1  ,\
+1 ,\
+1  ,\
0 \
\bigg{]}
\ ;
\end{equation}
\begin{equation}
a = -0.22 \ \hbox{ fm} \ .
\end{equation}
\begin{equation}
{{\rm K}^*{\rm N} }(0,{1\over 2})  =
{1\over 18} \ \bigg{[} \
-4  ,\
+5  ,\
-4  ,\
-3  \
\bigg{]}
\ ;
\end{equation}
\begin{equation}
a = +0.15 \ \hbox{ fm} \ .
\end{equation}
\vskip 0.5cm
\begin{equation}
{{\rm K}^*\Delta}(2,{5\over 2})  =
{1\over 3} \ \bigg{[} \
+{1}  ,\
-{1} ,\
-1  ,\
+{1} \
\bigg{]}
\ ;
\end{equation}
\begin{equation}
a = +0.14 \ \hbox{ fm} \ .
\end{equation}
\begin{equation}
{{\rm K}^*\Delta}(2,{3\over 2})  =
{1\over 18} \ \bigg{[} \
+{11}  ,\
-{1} ,\
-{1}  ,\
-{9} \
\bigg{]}
\ ;
\end{equation}
\begin{equation}
a = -0.20 \ \hbox{ fm} \ .
\end{equation}
\begin{equation}
{{\rm K}^*\Delta}(2,{1\over 2})  =
{1\over 9} \ \bigg{[} \
+{7}  ,\
+{1} ,\
+{1}  ,\
+{3} \
\bigg{]}
\ ;
\end{equation}
\begin{equation}
a = -0.86 \ \hbox{ fm} \ .
\end{equation}
\vskip 0.5cm
\begin{equation}
{{\rm K}^*\Delta}(1,S_{tot})  = -{1\over 3} \
{{\rm K}^*\Delta}(2,S_{tot})  \ \ \ \ \ \forall \ S_{tot}
\ ;
\end{equation}
\begin{equation}
{{\rm K}^*\Delta}(1,{5\over 2})  =
{1\over 9} \ \bigg{[} \
-{1}  ,\
+{1} ,\
+1  ,\
-{1} \
\bigg{]}
\ ;
\end{equation}
\begin{equation}
a = -0.05 \ \hbox{ fm} \ .
\end{equation}
\begin{equation}
{{\rm K}^*\Delta}(1,{3\over 2})  =
{1\over 54} \ \bigg{[} \
-{11}  ,\
+{1} ,\
+{1}  ,\
+{9} \
\bigg{]}
\ ;
\end{equation}
\begin{equation}
a = +0.07 \ \hbox{ fm} \ .
\end{equation}
\begin{equation}
{{\rm K}^*\Delta}(1,{1\over 2})  =
{1\over 27} \ \bigg{[} \
-{7}  ,\
-{1} ,\
-{1}  ,\
-{3} \
\bigg{]}
\ ;
\end{equation}
\begin{equation}
a = +0.29 \ \hbox{ fm} \ .
\end{equation}
Evidently attractive forces arise from the OGE spin-spin interaction in the
minimum-spin, minimum-isospin channels,

\begin{displaymath}
{\rm K}\Delta : \ \ \
(1,{3\over 2}) \ ;
\end{displaymath}

\begin{displaymath}
{\rm K}^* {\rm N}
: \ \ \
(0,{1\over 2}) \ ;
\end{displaymath}

\begin{displaymath}
{\rm K}^* \Delta
: \ \ \
(1,{1\over 2}) \ .
\end{displaymath}
The two exceptions to this rule are the K$^*\Delta$ channels
\begin{displaymath}
{\rm K}^* \Delta
: \ \ \
(2,{5\over 2})
\end{displaymath}
and
\begin{displaymath}
{\rm K}^* \Delta
: \ \ \
(1,{3\over 2}) \ ;
\end{displaymath}
although their weights sum to zero, variations in the detailed overlap
integrals lead to attractive OGE-hyperfine forces in these two channels as
well.

For our reference parameter set we find no molecular bound states; the
attractive forces are too weak to induce binding.
The experimental situation at present is rather confused; some references claim
evidence for resonances in several channels (see for example
\cite{HARW,Hash,Hyslop}), whereas other references
such as \cite{MO} and \cite{Watts} conclude that the same
phase shifts are nonresonant. Our results
do not support the most recent claims of resonances \cite{HARW}, since the
S-wave quantum numbers of our attractive channels do not correspond to those of
the negative parity candidates [$Z_0(1788),{3\over 2}^-$] and
[$Z_1(2074),{5\over 2}^-$].
However our negative result may be an
artifact of our approximations, including the neglect of spin-orbit effects and
couplings between channels. The spin-orbit effects are known from experiment to
be very important, and might be sufficient to lead to
Z$^*$-molecule bound states or strong
threshold enhancements in the attractive channels. Our negative result is
based on strong assumptions on the form of the interaction; this
should be relaxed in future theoretical work, and should not be used to argue
against experimental searches for possible Z$^*$ meson-baryon molecules.

\section{Summary and conclusions}

In this paper we have applied the quark Born diagram formalism to KN
scattering. In this approach one calculates hadron-hadron scattering amplitudes
in the nonrelativistic quark potential model assuming that the amplitude is the
coherent sum of all OGE interactions followed by all allowed quark line
exchanges; this is expected to be a useful description of reactions which are
free of $q\bar q$ annihilation. The model has few parameters, here
$\alpha_s/m_q^2$, $\rho=m_q/m_s$ and the hadron wavefunction parameters, and
with Gaussian wavefunctions the scattering amplitudes can be derived
analytically. The model was previously applied to I=2 $\pi\pi$ and I=3/2 K$\pi$
scattering with good results.

KN scattering is an important test of this approach because it is also
annihilation-free (at the valence quark level) and the meson and baryon
wavefunction parameters and the interaction strength are already reasonably
well established. Thus there is little freedom to adjust parameters. We find
good agreement with the experimental low energy I=0 and I=1 phase shifts given
standard quark model parameters. (The experimental I=0 scattering length is
usually claimed to be very small; we disagree with this interpretation of the
data and argue in support of a larger value.) A resolution of the disagreements
between different I=0 KN phase shift analyses, especially at very low energies,
is an important task for future experimental work. Hyslop \cite{Hyslop} also
suggests additional experimental work on the I=0 KN system. At higher energies
we find that the single Gaussian S-wave phase shifts fall with energy more
quickly than experiment given standard quark model parameters; we attribute
most of this effect to departures of the hadron wavefunction from single
Gaussians at short distances, perhaps in response to the attractive color
hyperfine interaction. We have confirmed that a smaller hadronic length scale
(about 0.7 times the usual nonrelativistic quark potential model
scale) gives S-wave phase shifts which are in
good agreement with experiment at all energies.

We have investigated the possibility of Z$^*$-molecule meson-baryon bound
states by extending our calculations to all channels allowed for
K$\Delta$, K$^*$N and K$^*\Delta$. Although we do find attractive interactions
in certain channels, in no case is the corresponding interhadron potential
sufficiently strong to form a bound state. Of course this result may be an
artifact of our approximations, in particular the assumption of keeping only
the spin-spin color hyperfine term and the single channel approximation. The
effect of relaxing these approximations would be a very interesting topic
for future study.

There are additional effects in the L$>$0 KN system which are known to be
important experimentally, which are not incorporated in our calculations of
single channel color hyperfine matrix elements. The most important of these is
a very large spin-orbit force, which it has not been possible to explain as an
OGE interaction \cite{VTJ,MP}.
Both this spin-orbit interaction and the Z$^*$ candidates may
be strongly affected by coupled channel effects, which we plan to investigate
in future work. Since much is already known experimentally about the reactions
KN$\to$K$^*$N and KN$\to$K$\Delta$, it should be possible to test
predictions of the quark Born diagrams for these channel couplings using
existing data sets. Although one might expect OPE forces to be important in
coupling KN to inelastic channels, such as in I=1 KN$\to$K$^*$N, the OPE
contribution to this process has been found experimentally to be small near
threshold \cite{bland2}. Thus experiment suggests that interquark forces such
as OGE and the confining interaction may be more important
than meson exchange in coupling KN to inelastic channels. We plan to evaluate
these offdiagonal couplings in detail in a future study.

\acknowledgements

We acknowledge useful contributions from R.Arndt, W.Bugg, S.Capstick,
F.E.Close, G.Condo, H.Feshbach, N.Isgur, R.Koniuk, M.D.Kovarik, K.Maltman,
B.R.Martin, D.Morgan, G.C.Oades, R.J.N.Phillips, B.Ratcliff, R.G.Roberts,
L.D.Roper, D.Ross,
S.Sorensen, J.Weinstein and R.Workman. This work was sponsored in
part by the United States Department of Energy under contracts
DE-AC02-76ER03069 with the Center for Theoretical Physics at the Massachusetts
Institute of Technology and DE-AC05-840R21400 with Martin Marietta Energy
Systems Inc.

\newpage

\section{appendix}

The procedure we advocate for describing hadronic interactions involves simply
calculating the Born order scattering amplitude for a given process using the
constituent quark model. In cases
where many channels contribute or one wishes to obtain nonperturbative
information (such as the possible existence of hadronic bound states) then one
must extract an effective potential (or effective potential matrix in the
case of multichannel problems) from the Born amplitude and integrate the
appropriate Schr\"odinger equation.
The validity of this procedure and its relationship to the resonating group
method are the subjects of this appendix.

Several theoretical complications arise when considering scattering of
composite
particles. For example, the Hamiltonian may be partitioned
in many different ways
corresponding to different rearrangement channels. Thus if $i$  represents
the initial channel consisting of hadrons $a$ and $b$, and $f$ represents the
final channel with hadrons $c$ and $d$, then we may write

\begin{equation}
H = H_i + V_i = H_f + V_f,
\end{equation}

\noindent
where

\begin{equation}
H_i =
-{1\over {2 \mu_{ab}} }
{\nabla_R^2 } + H_a + H_b
\end{equation}

\noindent
and

\begin{equation}
H_f = -{1\over {2 \mu_{cd}} }
{\nabla_R^2}+ H_c + H_d.
\end{equation}

\noindent
Here $\vec R$ is the appropriate
interhadron coordinate and $\mu_{ab}$ is the reduced mass
of the constituent masses of hadrons $a$ and $b$. The Hamiltonians
$H_a$ and $H_b$ describe the hadronic
wavefunctions in the initial channel. Thus we have

\begin{equation}
H_a \phi_{a,n}({\bf \xi_a}) = \epsilon_{a,n} \phi_{a,n}({\bf \xi_a})
\end{equation}

\noindent
and similarly for the other hadrons.

In general the wavefunction must be antisymmetrized appropriately and this
means summing over the various rearrangement channels with the correct weights.
Thus we take the incoming wavefunction to be $\phi_a({\bf \xi_a}) \phi_b({\bf
\xi_b}) \psi_0({\bf R})$ where $\psi_0$ is a plane wave and define the
antisymmetrization operator as

\begin{equation}
{\cal A} = \case{1}/{\sqrt{\eta}} \sum_P (-)^P P
\end{equation}
\noindent
where $P$ is a quark exchange operator.
The Born series for the process $ab \rightarrow cd$ may then be written as

\begin{equation}
\langle \hat f \vert t \vert \hat i \rangle = \case {1}/{\eta} \sum_{P P'}
(-)^P (-)^{P'} \langle P \phi_c \phi_d \psi_0
\vert \bigg( V_{P'} + V_P G_c V_{P'}
+ V_P G_c V_c G_c V_{P'} + \ldots \bigg)
\vert P' \phi_a \phi_b \psi_0 \rangle .
\label{borns}
\end{equation}

\noindent
Here $G_c$ is the Green function in a general channel $c$.

The Born order expression is

\begin{equation}
\langle \hat f \vert T \vert \hat i \rangle = \sum_P (-)^P \langle \phi_c
\phi_d \psi_0 \vert V_i \vert P \phi_a \phi_b \psi_0 \rangle.
\label{born}
\end{equation}

\noindent
This is the ``prior'' form of the T-matrix. The ``post'' form uses the
potential in the final channel rather than the initial channel. The expressions
are equivalent if the hadronic wavefunctions are exact and the T-matrix is
evaluated on energy shell. Note that these conditions must also hold if
the effective potential matrix is to be Hermitian.

Little is known about the convergence properties of the Born series. With no
exchange the conditions for convergence are probably similar to those for
simple potential scattering \cite{Joachain}. Of course exchange scattering
is necessarily present when describing hadronic interactions from the quark
level. At high energy
there is evidence that the lowest few Born terms can be accurate
\cite{Joachain}. However, since the potential is strong enough to cause
binding in the initial and final states, the series will diverge at low
energies. Nevertheless the strategy of extracting an effective potential
can be useful when the Born approximation is not accurate or even when it
diverges. We shall return to this point below.
Despite the theoretical problems, the small nuclear binding energy, the
small phase shifts seen in K$^+$N scattering, and the lack of quark model
state mixing evidenced in most of the meson spectrum all suggest the
utility of the Born approximation.

It should be stressed that the Born approximation can be useful even when
the effective potential is very strong. This will be true if the Born term
carries information on the dominant physics. Then the Born order scattering
amplitude may be Fourier transformed to yield an effective potential which
contains all of the dominant physics and may be integrated exactly. Since
hadronic interactions must involve constituents, this may not be carried out in
general, however it will be accurate if the new physics induced at higher
order in the Born series (such as polarization effects) does not dominate at
low energy. As will be discussed below, this appears to be true in many cases.

We now turn to the relationship of this approach to the resonating group
method. In the following we shall restrict our attention to the single
channel case. The resonating group Ansatz is then

\begin{equation}
\Psi = \phi_a({\bf \xi_a}) \phi_b ({\bf \xi_b}) \psi ({\bf R})
\end{equation}

\noindent
where $\psi$ is an unknown function of the interhadron distance. This Ansatz
must be antisymmetrized. For later convenience, we choose to separate the
identity permutation. Thus the wavefunction is

\begin{equation}
\hat \Psi = {\cal A} \Psi = \case{1}/{\sqrt{\eta}} \left(1 + \sum_{P\neq I}
(-)^P P \right) \phi_a \phi_b \psi.
\end{equation}

\noindent
Varying the  Schr\"odinger equation with respect to $\psi$ and
rewriting the resulting expression for $\psi$ in Lippmann-Schwinger form
yields the following equation.

\begin{displaymath}
\psi({\bf R}) = \psi_0({\bf R}) + 2 \mu_{ab} \int G_0({\bf R},{\bf R'})
V_D({\bf R'}) \psi({\bf R'}) d{\bf R'}
\end{displaymath}
\begin{displaymath}
- \sum_{P \neq I}  (-)^P \int G_0({\bf R},{\bf R'}) \phi_a^*({\bf \xi_a})
\phi_b^*({\bf \xi_b})\,  \left[ \nabla_R^2 + k_{\rm rel}^2 \right]
P[\phi_a({\bf \xi_a}) \phi_b({\bf \xi_b}) \psi({\bf R'})] d{\bf \xi_a}
d{\bf \xi_b} d{\bf R'}
\end{displaymath}
\begin{equation}
 + 2 \mu_{ab} \sum_{P \neq I}  (-)^P\int G_0({\bf R},{\bf R'})
\phi_a^*({\bf \xi_a})  \phi_b^*({\bf \xi_b}) \, V_i({\bf \xi_a},{\bf \xi_b},
{\bf R'}) \nonumber \\
P[\phi_a({\bf \xi_a}) \phi_b({\bf \xi_b}) \psi{\bf R'}) ]
 d{\bf \xi_a} d{\bf \xi_b} d{\bf R'}  \\
\label{LS}
\end{equation}

\noindent
where

\begin{equation}
k_{\rm rel}^2 = 2 \mu_{ab} (E - \epsilon_a - \epsilon_b)
\end{equation}

\noindent
and

\begin{equation}
V_D({\bf R}) = \int \phi_a^*({\bf \xi_a}) \phi_b^*({\bf \xi_b}) V_i({\bf
\xi_a},
{\bf \xi_b}, {\bf R}) \phi_a({\bf \xi_a}) \phi_b({\bf \xi_b}) d{\bf \xi_a}
d{\bf \xi_b},
\end{equation}

\noindent
and $G_0$ is the Green function for $\nabla_R^2 + k_{\rm rel}^2$.

The permutation operator in the third term implies that
$\nabla_R^2 + k_{\rm rel}^2$ is a Hermitian operator and may be safely applied
to the left. Thus the third term (the kinetic and energy exchange kernels)
simplifies to

\begin{equation}
-\sum_{P\neq I} (-)^P \int \phi_a \phi_b P[ \phi_a \phi_b \psi({\bf R})]
d{\bf \xi_a} d{\bf \xi_b} \equiv \int N({\bf R},{\bf \tilde R}) \psi({\bf
\tilde R}) d{\bf \tilde R}
\end{equation}

\noindent
where $N({\bf R}, {\bf \tilde R})$ is the normalization kernel.
Because of nontrivial permutation operators this expression is damped as
$R \rightarrow \infty$ and hence it does not contribute to scattering.

We may now iterate Eq.\ (\ref{LS}) to see that it corresponds to the full Born
series (\ref{borns}) with the sum over intermediate states restricted to the
appropriate single channel. In particular the $R \rightarrow \infty$ limit of
the first term in the series corresponds to the Born order T-matrix of Eq.\
(\ref{born}) (for the case of elastic scattering).

Eq.\ ({\ref{LS}) indicates that setting $V_i = 0$ in a resonating group
calculation should yield a null phase shift. However if one uses approximate
hadronic wavefunctions (as is almost always the case in resonating group
calculations) then a residual spurious phase shift will remain. We note that
Bender {\it et al.} \cite{BDPK} employ single Gaussian hadronic wavefunctions
so that one expects small phase shifts upon setting $V_i = 0$. This is
indeed what they found for I=0 K$^+$N scattering. However they obtained
rather large phase shifts for the I=1 case. Since the Hamiltonian is
independent of the isospin the Gaussian wavefunctions should have been
equally effective in both cases and one must conclude that there is
likely to be an error in the I=1 calculation. Maltman \cite{Kim}
has concluded that there
are indeed errors in the hyperfine matrix elements in this reference.

Solving the single channel resonating group equation is similar to the process
of extracting an effective potential from the Born scattering amplitude and
integrating it exactly. Both methods treat the single channel subspace
nonperturbatively and hence are successful when the single channel
approximation is a good one. Both methods fail if (off-channel) virtual
particles, polarization, wavefunction distortion, and similar effects dominate
the low energy behavior of the system. As stated above, this does not seem to
happen in practice. The similarity of hadron scattering amplitudes obtained
from the single channel resonating group method and from integrating effective
potentials has been demonstrated for several cases in Ref. \cite{Swan}.

\end{document}